\documentclass[twocolumn,aps,pra,floatfix,amsmath,amssymb,showpacs,nofootinbib]{revtex4} 
\usepackage{graphicx}

\newcommand{\ket}[1]{\ensuremath{|{#1\rangle}}}

\newcommand{\ketbra}[2]{\ensuremath{|{#1 \rangle}{\langle #2}|}}
\newcommand{\E}{\ensuremath{\text{e}}}
\newcommand{\e}{\ensuremath{\text{e}}}
\newcommand{\D}{\ensuremath{\text{d}}}
\newcommand{\I}{\ensuremath{\text{i}}}
\newcommand{\op}[1]{\hat{#1}}

\begin{document}

\title{Decoherence and dissipation of a quantum harmonic
  oscillator\\coupled to two-level systems}

\author{Maximilian Schlosshauer}
\email{m.schlosshauer@unimelb.edu.au}
\affiliation{School of Physics, The University of Melbourne,
  Melbourne, Victoria 3010, Australia}
  
  \author{Andrew P.\ Hines}

  \affiliation{Pacific Institute of Theoretical Physics, Department of
    Physics and Astronomy, University of British Columbia, Vancouver
    BC, Canada V6T 1Z1}

\author{Gerard J.\ Milburn}

\affiliation{School of Physical Sciences, The University of
  Queensland, Brisbane, Queensland 4072, Australia}

\date{\today}

\begin{abstract}
  We derive and analyze the Born--Markov master equation for a quantum
  harmonic oscillator interacting with a bath of independent two-level
  systems. This hitherto virtually unexplored model plays a
  fundamental role as one of the four ``canonical''
  system--environment models for decoherence and dissipation. To
  investigate the influence of further couplings of the environmental
  spins to a dissipative bath, we also derive the master equation for
  a harmonic oscillator interacting with a single spin coupled to a
  bosonic bath. Our models are experimentally motivated by
  quantum-electromechanical systems and micron-scale ion traps.
  Decoherence and dissipation rates are found to exhibit temperature
  dependencies significantly different from those in quantum Brownian
  motion. In particular, the systematic dissipation rate for the
  central oscillator \emph{decreases} with increasing temperature and
  goes to zero at zero temperature, but there also exists a
  temperature-independent momentum-diffusion (heating) rate.
\end{abstract}

\pacs{03.65.Yz, 42.50.Lc, 85.85.+j, 37.10.Ty}

\maketitle

\section{Introduction}

\begin{figure}
 \includegraphics[scale=.85]{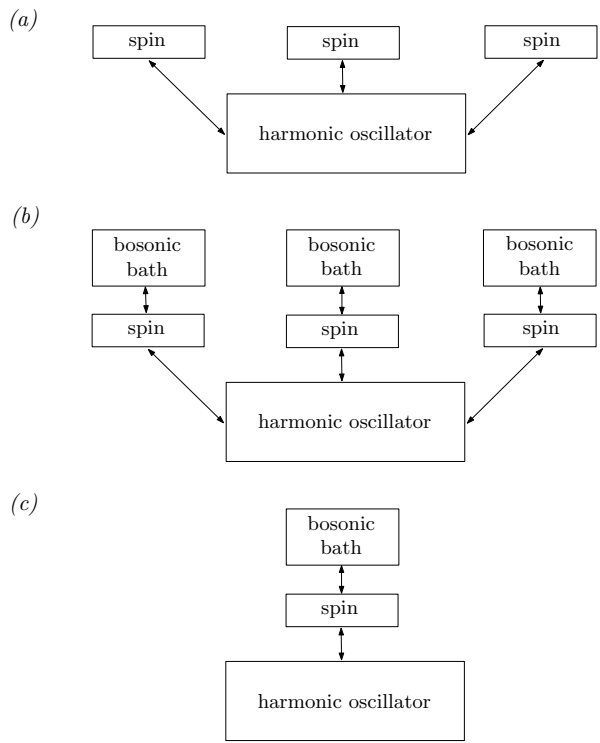}
 \caption{\label{fig:sb}\emph{(a)} Oscillator--spin model. \emph{(b)}
   Oscillator--spin model with each spin coupled to an additional
   bosonic bath. \emph{(c)} Limiting case of \emph{(b)} in which the
   oscillator interacts with a single environmental spin coupled to a
   bosonic bath.}
\end{figure}

Theoretical studies of decoherence and dissipation in quantum systems
have hitherto focused on three \emph{canonical} system--environment
models: (i) A harmonic oscillator (or, more generally, a particle
moving in phase space) coupled to a bath of other harmonic oscillators
(quantum Brownian motion)
\cite{Hu:1992:om,Grabert:1988:bf,Caldeira:1983:on,Haake:1932:tt}; (ii)
a quantum two-level system (TLS), represented by a spin-$\frac{1}{2}$
particle, interacting with a bath of harmonic oscillators (spin--boson
model) \cite{Leggett:1987:pm}; and (iii) a spin-$\frac{1}{2}$ particle
coupled to a bath of other spins (spin--spin model)
\cite{Prokofev:2000:zz}. Surprisingly, however, the fourth possible
canonical combination, namely, a single harmonic oscillator
interacting with a bath of spin-$\frac{1}{2}$ particles---which, in
obvious nomenclature, shall henceforth be referred to as the
\emph{oscillator--spin model} (Fig.~\ref{fig:sb}a)---has not yet been
studied in any detail in the literature in terms of a Markovian
master equation.

It is the purpose of this paper to close this gap by giving a
microscopic treatment of the oscillator--spin model. We will derive
the Born--Markov master equation and compare the resulting dynamics to
those induced by an oscillator bath (quantum Brownian motion). Apart
from its relevance in completing the set of canonical models, the
oscillator--spin model is also motivated by recent experiments on
quantum-electromechanical systems (QEMS)
\cite{Blencowe:2004:mm,Schwab:2005:hb,Roukes:2001:uu} and micron-scale
ion traps. In both systems, a central quantum-mechanical vibrational
degree of freedom interacts with two-level defects causing dissipation
and decoherence of the oscillator. We may represent this situation by
a harmonic oscillator coupled to a collection of TLS, i.e., by a model
of the oscillator--spin type.

QEMS are nanometer-to-micrometer--sized crystalline mechanical
resonators coupled to nanoscale electronic transducers that detect the
high-frequency (MHz--GHz) vibrational motion of the resonator. Since
only the lowest, fundamental flexural mode of the resonator turns out
to be relevant \cite{Blencowe:2004:mm}, the resonator can be modeled
as a single quantum-mechanical harmonic oscillator. Recent
experimental evidence \cite{Mohanty:2002:mm,Zolfagharkhani:2005:tv}
(see also the molecular-dynamics simulation of \cite{Chu:2007:pm} and
earlier results in \cite{Kleiman:1987:ff,Mihailovich:1992:yy})
strongly suggests that the dominant source of decoherence and
dissipation in QEMS is the interaction with two-level defects
intrinsically present in the resonator itself.

In ion traps a single ion can be trapped by a time-dependent potential
and cooled to very low energies \cite{Leibfried:2003:om}. Under the
right conditions the motion of the ion is equivalent to that of a
quasi-one-dimensional harmonically bound particle. A major source of
decoherence in ion traps is thought to arise from fluctuating patch
potentials on the trap electrodes \cite{Turchette:2000:oa}. Roughly
speaking, this causes a fluctuating linear potential that results in
random forces acting on the ion. The net effect is a slow heating of
the ions.  This problem is particularly acute in small traps
\cite{Stick:2006:aa,Seidelin:2006:rz}, where anomalous heating has
been experimentally observed \cite{Deslauriers:2006:mz}. More recent
experiments have cooled the traps to a few~K and seen a dramatic
reduction in heating \cite{Labaziewicz:2007:im}. We anticipate that
with further cooling the heating will ultimate be attributable to
charge fluctuations in two-level traps, especially for oxide barriers
in semiconductor substrates.

In many cases, it is reasonable to assume that each of the two-level
defects will be also coupled to its own environment, which we may
model as a bosonic bath. We are thus led to a more complicated model
in which the central oscillator couples to a collection of independent
spin--boson models (Fig.~\ref{fig:sb}b). In this paper we will
consider the special case of only a single TLS interacting with a
bosonic bath (Fig.~\ref{fig:sb}c). This simplification allows us to
analytically derive the master equation for the central oscillator in
the limit in which the oscillator is strongly coupled to a
steady-state TLS. It is also experimentally motivated by the fact that
in GHz QEMS and micron-scale, cryogenic ion traps the number of
defects that participate in the dynamics is thought to be quite small.

We emphasize that the focus of this paper is a study of the general
dynamics and properties of the oscillator--spin model in the context
of the canonical models, and it is not our aim to present detailed
models for ion traps and QEMS. However, these systems lend urgent
experimental relevance to the oscillator--spin model, and our model
may serve as a starting point for the development of models tailored
to specific experimental situations.  In existing models of ion traps,
the fluctuating forces have thus far been treated classically
\cite{Savard:1997:um,James:1998:za,Schneider:1998:yz,%
  Grotz:2006:km,Brouard:2004:in}.  For QEMS, a realistic and
quantitatively accurate modeling of the influence of the various
defects on the resonator is rather involved. First theoretical studies
(see, e.g., \cite{Ahn:2003:mt,Zolfagharkhani:2005:tv}) were recently
followed by detailed work by Seo\'anez, Guinea, and Castro Neto
\cite{Seoanez:2006:yb,Seoanez:2007:um}.  In order to be able to use a
spectral-function treatment for the environmental TLS, these authors
focused on the limit $k_\text{B} T \gg \hbar \Omega_0$, where
$\Omega_0$ is the natural frequency of the resonator. However,
attaining the quantum regime of QEMS requires the opposite limit
$k_\text{B} T \ll \hbar \Omega_0$ (GHz QEMS).  In this case, it is
likely that only very few TLS will be relevant to the dynamics of the
resonator and we suggest that the correct description may be closer to
the model discussed in Sec.~\ref{sec:oscill-coupl-single} below.

This paper is organized as follows. Sec.~\ref{sec:derivation} presents
the derivation of the Born--Markov master equation for the central
oscillator coupled to a spin bath. In
Sec.~\ref{sec:oscill-coupl-single} we derive the master equation for a
harmonic oscillator interacting with a single spin coupled to a
bosonic bath. We summarize our results in Sec.~\ref{sec:conclusions}.

\section{\label{sec:derivation}Master equation for a
  harmonic oscillator coupled to a spin bath}

\subsection{Model}

We consider a single quantum harmonic oscillator (the system
$\mathcal{S}$) with self-Hamiltonian
\begin{equation}
  \label{eq:7jd}
  \op{H}_\mathcal{S} =  \frac{\op{P}^2}{2M} + \frac{M\Omega_0}{2} \op{X}^2.
\end{equation}
The oscillator interacts with an environment $\mathcal{E}$ of $N$
independent spin-$\frac{1}{2}$ particles (quantum TLS). The
environment is described by the self-Hamiltonian (setting $\hbar
\equiv 1$)
\begin{equation}
  \label{eq:7ahsdg}
  \op{H}_\mathcal{E} \equiv \sum_{i=1}^N \op{H}^{(i)}_\mathcal{E} = \sum_{i=1}^N \frac{\omega_i}{2}
  \op{\sigma}_z^{(i)} +  \sum_{i=1}^N
  \frac{\Delta_i}{2} \op{\sigma}_x^{(i)},
\end{equation}
where $\omega_i$ and $\Delta_i$ are, respectively, the asymmetry
energy and tunneling matrix element of the $i$th bath spin. The environment
couples linearly to the position coordinate of the oscillator via the
interaction Hamiltonian
\begin{equation}
  \label{eq:ssd9}
  \op{H}_\text{int}  = \op{X} \otimes \sum_{i=1}^N g_i
  \op{\sigma}_z^{(i)} \equiv \op{X} \otimes \op{E}. 
\end{equation}
The total system--environment combination is then described by the
Hamiltonian
\begin{equation}
  \label{eq:sfksf1}
  \op{H} = \op{H}_\mathcal{S} +\op{H}_\mathcal{E} + \op{H}_\text{int}. 
\end{equation}
We assume the limit of weak system--environment couplings and no
initial system--environment correlations, $\op{\rho}(0) =
\op{\rho}_\mathcal{S}(0) \otimes \op{\rho}_\mathcal{E}(0)$. We take
the environment to be in thermal equilibrium at temperature $T$. Since
the spins of this thermal bath are independent, we have
\begin{equation}
  \label{eq:5}
  \op{\rho}_\mathcal{E}(0) = \frac{1}{Z}\e^{- \beta \op{H}_\mathcal{E} }
  \equiv \frac{1}{Z} \prod_{i=1}^N \e^{-\beta
    \op{H}^{(i)}_\mathcal{E}}, 
\end{equation}
where $\beta\equiv 1/k_\text{B}T$ and $Z = \text{Tr}_\mathcal{E} \,
\e^{- \beta \op{H}_\mathcal{E} }$.  We would now like to derive the
Born--Markov master equation for this spin-bath model. For the
interaction Hamiltonian \eqref{eq:ssd9}, the general form of the
master equation is \cite{Schlosshauer:2007:un}
\begin{align}
\label{eq:born-markov-master}
\frac{\D}{\D t} \op{\rho}_\mathcal{S}(t) &= -\I \left[
  \op{H}_\mathcal{S}, \op{\rho}_\mathcal{S}(t) \right] \notag \\ &\quad -
\int_0^\infty \D \tau \, \biggl\{ \mathcal{C}(\tau) \left[
    \op{X}, \op{X}(-\tau) \op{\rho}_\mathcal{S}(t) \right] \notag \\
  & \qquad\qquad + \mathcal{C}(-\tau)\left[ \op{\rho}_\mathcal{S}(t)\op{X}(-\tau),
    \op{X} \right] \biggr\}.
\end{align}
Here 
\begin{equation}
\label{eq:fuis764w4}
  \op{X}(\tau) = 
  \op{X} \cos \left( \Omega_0 \tau \right) + \frac{1}{M\Omega_0} \op{P}
  \sin \left( \Omega_0 \tau \right)
\end{equation}
denotes the system's position operator $\op{X}$ in the interaction
picture. The spin-environment self-correlation function
$\mathcal{C}(\tau)$ appearing in Eq.~\eqref{eq:born-markov-master} is
given by
\begin{equation}
  \label{eq:pmpnnun12}
 \mathcal{C}(\tau) \equiv \left\langle \op{E}(\tau) \op{E}
  \right\rangle_{\op{\rho}_\mathcal{E}},
\end{equation}
where $\op{E}(\tau)= \E^{\I \op{H}_\mathcal{E} \tau }\op{E}\E^{-\I
  \op{H}_\mathcal{E} \tau} $ is the environment operator $\op{E}$ in
the interaction picture and the average is taken over the initial
state $\op{\rho}_\mathcal{E}\equiv \op{\rho}_\mathcal{E}(0)$ of the
environment (the Born approximation means that $\op{\rho}(t) \approx
\op{\rho}_\mathcal{S}(t) \otimes \op{\rho}_\mathcal{E}(0)$ for all
$t$).

\subsection{Calculation of the environment self-correlation function}

First, we compute the environment self-correlation function
\eqref{eq:pmpnnun12}, which we may write as 
\begin{align}
  \label{eq:kjJbd6FUT22}
   \mathcal{C}(\tau) &=  \sum_{ij} g_ig_j \left\langle \E^{\I
    \op{H}^{(i)}_\mathcal{E} \tau} \op{\sigma}_z^{(i)}
 \E^{- \I \op{H}^{(i)}_\mathcal{E} \tau}
  \op{\sigma}_z^{(j)} \right\rangle_{\op{\rho}_\mathcal{E}} \notag \\ &\equiv
\sum_{ij} g_ig_j \left\langle \op{\sigma}_z^{(i)}(\tau)
  \op{\sigma}_z^{(j)} \right\rangle_{\op{\rho}_\mathcal{E}}.
\end{align}
Because the environmental spins do not directly interact with each
other, they are uncorrelated,
\begin{equation}
  \label{eq:fsdlnjHIGI24}
  \left\langle \op{\sigma}_z^{(i)}(\tau) 
    \op{\sigma}_z^{(j)} \right\rangle_{\op{\rho}_\mathcal{E}} =
  \left\langle \op{\sigma}_z^{(i)}(\tau)
  \right\rangle_{\op{\rho}_\mathcal{E}}   \left\langle
    \op{\sigma}_z^{(j)} \right\rangle_{\op{\rho}_\mathcal{E}} \quad \text{for $i\not=j$},
\end{equation}
and thus Eq.~\eqref{eq:kjJbd6FUT22} can be decomposed as
\begin{multline}
\label{eq:fsbvnHUH29}
\mathcal{C}(\tau) =\sum_i g_i \left\langle\op{\sigma}^{(i)}_z(\tau)
  \right\rangle_{\op{\rho}_\mathcal{E}} \sum_{j\not=i} g_j \left\langle\op{\sigma}^{(j)}_z
  \right\rangle_{\op{\rho}_\mathcal{E}} \\ + \sum_{i} g_i^2
  \left\langle\op{\sigma}^{(i)}_z(\tau) \op{\sigma}^{(i)}_z
  \right\rangle_{\op{\rho}_\mathcal{E}}.
\end{multline}
Let us assume that at $t=0$ the average of the ``quantum force'' due
to the collective action of all environmental spins is equal to zero,
\begin{equation}
\label{eq:kjJbAAd6FsdfsUT22}
  \left\langle \op{E} \right\rangle_{\op{\rho}_\mathcal{E}} = \sum_i
  g_i \left\langle\op{\sigma}^{(i)}_z \right\rangle_{\op{\rho}_\mathcal{E}} = 0.
\end{equation}
This is a nonrestrictive assumption, since
Eq.~\eqref{eq:kjJbAAd6FsdfsUT22} can always be fulfilled by simply
adding a constant to the Hamiltonian. Then the term $\sum_{j\not=i}
g_j \left\langle\op{\sigma}^{(j)}_z
\right\rangle_{\op{\rho}_\mathcal{E}}$ appearing in
Eq.~\eqref{eq:fsbvnHUH29} will also tend to zero, and
Eq.~\eqref{eq:kjJbd6FUT22} simplifies to
\begin{align}
  \label{eq:kjJbAAd6FUT22}
  \mathcal{C}(\tau) &= \sum_i g_i^2 \left\langle
    \op{\sigma}_z^{(i)}(\tau)
    \op{\sigma}_z^{(i)} \right\rangle_{\op{\rho}_\mathcal{E}} \notag \\
  &= \sum_i g_i^2 \, \text{Tr}_\mathcal{E} \left\{ \left[ \frac{1}{Z}
      \prod_i \E^{ - \op{H}_\mathcal{E}^{(i)} /k_\text{B}T} \right]
    \op{\sigma}_z^{(i)} (\tau)  \op{\sigma}_z^{(i)} \right\} \notag \\
  &= \sum_i g_i^2 \frac{1}{Z_i} \text{Tr}_{\mathcal{E}_i} \left\{ \E^{
      - \op{H}^{(i)}_\mathcal{E} /k_\text{B}T} \op{\sigma}_z^{(i)}
    (\tau) \op{\sigma}_z^{(i)} \right\},
\end{align}
where $Z_i = \text{Tr}_{\mathcal{E}_i} \, \E^{-
  \op{H}_{\mathcal{E}_i}/ k_\text{B}T} $ and in the last line we have
again used the fact that the bath spins are uncorrelated.

To calculate the interaction-picture operator
$\op{\sigma}_z^{(i)}(\tau) = \E^{- \I \op{H}^{(i)}_\mathcal{E} \tau}
\op{\sigma}_z^{(i)} \E^{- \I \op{H}^{(i)}_\mathcal{E} \tau}$, we write
the environment Hamiltonian $\op{H}^{(i)}_\mathcal{E}$ in matrix form
in the eigenbasis $\{ \ket{0}_i, \ket{1}_i \}$ of
$\op{\sigma}_z^{(i)}$,
\begin{align}
  \op{H}_\mathcal{E}^{(i)} = \frac{1}{2} \left( \begin{array}{cc}
      \omega_i & \Delta_i  \\
      \Delta_i & - \omega_i
\end{array} \right).
\end{align}
The matrix eigenvalues are $E_\pm^{(i)} = \pm\frac{1}{2} \sqrt{
  \omega_i^2 + \Delta_i^2 } \equiv \pm \frac{1}{2}
\widetilde{\omega}_i$ with corresponding eigenvectors
\begin{subequations}
\begin{align}
  \ket{+}_i &= \cos \frac{\theta_i}{2} \ket{0}_i + \sin \frac{\theta_i}{2} \ket{1}_i,  \\
  \ket{-}_i &= -\sin \frac{\theta_i}{2}\ket{0}_i + \cos
  \frac{\theta_i}{2} \ket{1}_i,
\end{align}
\end{subequations}
where $\theta_i = \arctan \frac{\Delta_i}{\omega_i}$. With respect to
the basis $\{\ket{+}_i, \ket{-}_i \}$ , the matrix representation of
$\op{\sigma}_z^{(i)}$ reads
\begin{align}
  \op{\sigma}_z^{(i)} = \left( \begin{array}{cc}
      \cos\theta_i & -\sin\theta_i  \\
      -\sin\theta_i & - \cos\theta_i
\end{array} \right).
\end{align}
We can now evaluate Eq.~\eqref{eq:kjJbAAd6FUT22} directly by carrying
out the relevant matrix products and then taking the trace. The result
is
\begin{multline}
\label{eq:4}
\mathcal{C}(\tau) = \mathcal{C}_0 + \sum_{i}
\left(\frac{g_i\Delta_i}{\widetilde{\omega}_i}\right)^2 \bigl[
  \cos\left(\widetilde{\omega}_i\tau\right)\\ - \I
  \tanh(\beta\widetilde{\omega}_i/2)
  \sin\left(\widetilde{\omega}_i\tau\right) \bigr],
\end{multline}
where $\mathcal{C}_0 = \sum_{i}
\left(\frac{g_i\omega_i}{\widetilde{\omega}_i}\right)^2$ is a
time-independent constant.

\subsection{Continuum limit}

Let us introduce the spectral density function
\begin{equation}
\label{eq:vdfpmdBGDJmv16as}
J(\widetilde{\omega}) \equiv \sum_i  \left(\frac{g_i\Delta_i}{\widetilde{\omega}_i}\right)^2
\delta(\widetilde{\omega}-\widetilde{\omega}_i) 
\end{equation}
and write Eq.~\eqref{eq:4} as
\begin{align}
\label{eq:89ydhgvjkdhsjgk4}
\mathcal{C}(\tau) &= \mathcal{C}_0 + \int_0^\infty \D
\widetilde{\omega} \, J(\widetilde{\omega}) \bigl[
\cos\left(\widetilde{\omega}\tau\right) \notag \\ &\qquad\qquad - \I
\tanh(\beta \widetilde{\omega}/2)
\sin\left(\widetilde{\omega}\tau\right) \bigr] \notag \\ &\equiv
\mathcal{C}_0 + \nu(\tau) - \I \eta(\tau).
\end{align}
The functions $\nu(\tau)$ and $\eta(\tau)$ take the same functional
form as the noise and dissipation kernels, respectively, in the case
of an oscillator bath (quantum Brownian motion) with spectral density
$J_\text{osc}(\widetilde{\omega})$,
\begin{subequations}
\label{eq:72y3fqufeq1}
\begin{align}
  \nu_\text{osc}(\tau) &= \int_0^\infty \D \widetilde{\omega} \,
  J_\text{osc}(\widetilde{\omega}) \coth
  \left(\frac{\widetilde{\omega}}{2k_\text{B} T}\right) \cos
  \left(\widetilde{\omega}\tau\right), \label{eq:vdjpoo17} \\
  \eta_\text{osc}(\tau) &= \int_0^\infty \D \widetilde{\omega}\,
  J_\text{osc}(\widetilde{\omega})
  \sin\left(\widetilde{\omega}\tau\right), \label{eq:ponol218}
\end{align}
\end{subequations}
\emph{provided} we choose
\begin{equation}
J(\widetilde{\omega}) =
J_\text{osc}(\widetilde{\omega})\coth\left(\frac{\widetilde{\omega}}{2k_\text{B}T}\right)
\end{equation}
in Eq.~\eqref{eq:89ydhgvjkdhsjgk4}.  Conversely, ignoring the constant
term $\mathcal{C}_0$, we can map the spin bath with spectral density
$J(\widetilde{\omega})$ onto an oscillator bath with ``surrogate''
spectral density
\begin{equation}
\label{eq:89542yhjfgkh2}
J_\text{osc}(\widetilde{\omega}) = J(\widetilde{\omega})
\tanh\left(\frac{\widetilde{\omega}}{2k_\text{B}T}\right). 
\end{equation}
This is an example of the general result, first derived by Feynman and
Vernon \cite{Feynman:1963:jj}, that in the limit of sufficiently weak
coupling \emph{any} dissipative bath (including the spin bath) can be
mapped onto a bath of oscillators. We note that in the limit $\Delta_i
\gg \omega_i$ for all $i$ (and thus $\mathcal{C}_0 \rightarrow 0$),
expression \eqref{eq:89ydhgvjkdhsjgk4} coincides with a result
previously obtained by Caldeira, Castro Neto, and de Carvalho
\cite{Caldeira:1993:bz} for a model of a general system interacting
linearly and weakly with a spin bath, where the environmental
self-Hamiltonian was assumed to take a more simple form than in our
model.

\subsection{Master equation}

Inserting Eq.~\eqref{eq:89ydhgvjkdhsjgk4} into
Eq.~\eqref{eq:born-markov-master} and using Eq.~\eqref{eq:fuis764w4}
leads to the master equation
\begin{align}
\label{eq:vfoinbn76fGGd9s27}
\frac{\D}{\D t} \op{\rho}_\mathcal{S}(t) &= -\I \left[
  \op{H}_\mathcal{S} + \frac{1}{2}M \widetilde{\Omega}_0^2 \op{X}^2,
  \op{\rho}_\mathcal{S}(t) \right] \notag \\ & \quad - \I \gamma
\left[ \op{X}, \left\{ \op{P}, \op{\rho}_\mathcal{S}(t) \right\}
\right] - D \left[ \op{X}, \left[ \op{X}, \op{\rho}_\mathcal{S}(t)
  \right] \right] \notag \\ & \quad- f \left[ \op{X}, \left[ \op{P},
    \op{\rho}_\mathcal{S}(t) \right] \right].
\end{align}
Here, the coefficients $\widetilde{\Omega}_0^2$, $\gamma$, $D$, and $f$
are defined as
\begin{subequations}\label{eq:jcsfr09355378}
\begin{align}
  \widetilde{\Omega}_0^2 &\equiv - \frac{2}{M} \int_0^\infty \D \tau
  \,
  \eta(\tau) \cos\left( \Omega_0 \tau \right), \label{eq:caytcs1} \\
  \gamma &\equiv \frac{1}{M\Omega_0} \int_0^\infty \D \tau \,
  \eta(\tau) \sin\left( \Omega_0 \tau \right), \label{eq:caytcs2} \\
  D &\equiv \int_0^\infty \D \tau \, \left[\mathcal{C}_0 + \nu(\tau)
  \right] \cos\left( \Omega_0 \tau \right) \notag \\ &\equiv D_0 +
  \int_0^\infty \D \tau \, \nu(\tau) \cos\left( \Omega_0 \tau
  \right) \notag \\ &\equiv D_0 + D_1, \label{eq:caytcs3}  \\
  f &\equiv - \frac{1}{M\Omega_0} \int_0^\infty \D \tau \, \left[
    \mathcal{C}_0 + \nu(\tau) \right] \sin\left( \Omega_0 \tau \right)
  \notag \\ &\equiv f_0 - \frac{1}{M\Omega_0} \int_0^\infty \D \tau \,
  \nu(\tau) \sin\left( \Omega_0 \tau \right)\notag \\ & \equiv
  f_0+f_1. \label{eq:caytcs4}
\end{align}
\end{subequations}
The interpretation of these coefficients is analogous to the case of
quantum Brownian motion. The coefficient $\widetilde{\Omega}_0^2$
describes a frequency shift (``Lamb-shift'' renormalization of the
natural frequency of the oscillator), $\gamma$ is the momentum-damping
(and thus dissipation) rate, and $D$ and $f$ are the normal-diffusion
and anomalous-diffusion coefficients describing decoherence.  We see
that $\widetilde{\Omega}_0^2$ and $\gamma$ are explicitly
temperature-dependent while $D$ and $f$ are not, which is exactly
opposite as in quantum Brownian motion. Formally, this difference is
easily understood from the fact that using the surrogate spectral
density \eqref{eq:89542yhjfgkh2} in the expressions for the
oscillator-bath noise and dissipation kernels \eqref{eq:72y3fqufeq1}
eliminates the temperature-dependent term in the integral
\eqref{eq:vdjpoo17} while introducing the term $\tanh
\left(\Omega_0/2k_\text{B} T\right)$ in the integral
\eqref{eq:ponol218}.

\subsection{Example: Ohmic spectral density}

Let us consider an ohmic spectral density for the spin bath with a
Lorentz--Drude high-frequency cutoff,
\begin{equation}
  \label{eq:AApojsvsddsjldfv1}
  J(\widetilde{\omega}) = \frac{2M\gamma_0}{\pi} \widetilde{\omega}
  \frac{\Lambda^2}{\Lambda^2 + \widetilde{\omega}^2}.
\end{equation}
The coefficient $\gamma$ is given by a double Fourier sine transform
of the function $J(\widetilde{\omega})\tanh(\beta
\widetilde{\omega}/2)$, which returns the original function up to a
prefactor of $\pi/2$,
\begin{align}
  \label{eq:poisjuw48yuKKHG2}
  \gamma &= \frac{1}{M\Omega_0} \int_0^\infty \D \tau \, \sin\left(
    \Omega_0 \tau \right) \notag \\ & \quad \times \int_0^\infty \D
  \widetilde{\omega} \, J(\widetilde{\omega}) \tanh(\beta
  \widetilde{\omega}/2) \sin\left(\widetilde{\omega}\tau\right) \notag
  \\ &= \gamma_0 \frac{\Lambda^2}{\Lambda^2 + \Omega_0^2} \tanh\left(
    \frac{\Omega_0}{2k_\text{B} T}\right).
\end{align}
Similarly, the coefficient $D_1$ is given by a double Fourier cosine
transform of $J(\widetilde{\omega})$, which leads to
\begin{equation}
\label{eq:pNUNjuw48yuKKHG2}
  D = D_0 + M\gamma_0 \Omega_0
  \frac{\Lambda^2}{\Lambda^2 + \Omega_0^2}.
\end{equation}
For quantum Brownian motion these coefficients read
\cite{Schlosshauer:2007:un}
\begin{subequations}
\begin{align}
  \label{eq:poisjuw48yu2}
  \gamma_\text{QBM} &=  \gamma_0  \frac{\Lambda^2}{\Lambda^2 + \Omega_0^2}, \\
  D_\text{QBM} &= M\gamma_0 \Omega_0 \frac{\Lambda^2}{\Lambda^2 +
    \Omega_0^2} \coth\left( \frac{\Omega_0}{2k_\text{B} T}\right).
  \label{eq:poisjuw48yudfkskjbkl2}
\end{align}
\end{subequations}
Disregarding the constant term $D_0$, we see that the spin-bath
coefficients are given by the oscillator-bath coefficients multiplied
by the term $\tanh \left(\Omega_0/2k_\text{B} T\right)$, which is
simply a direct consequence of the use of the surrogate spectral
density \eqref{eq:89542yhjfgkh2}.

\subsection{\label{sec:comp-betw-spin}Comparison between spin and
  oscillator baths}

\begin{figure}
\begin{flushleft} \emph{(a)} \end{flushleft}

\vspace{-0.5cm}

\includegraphics[scale=.3]{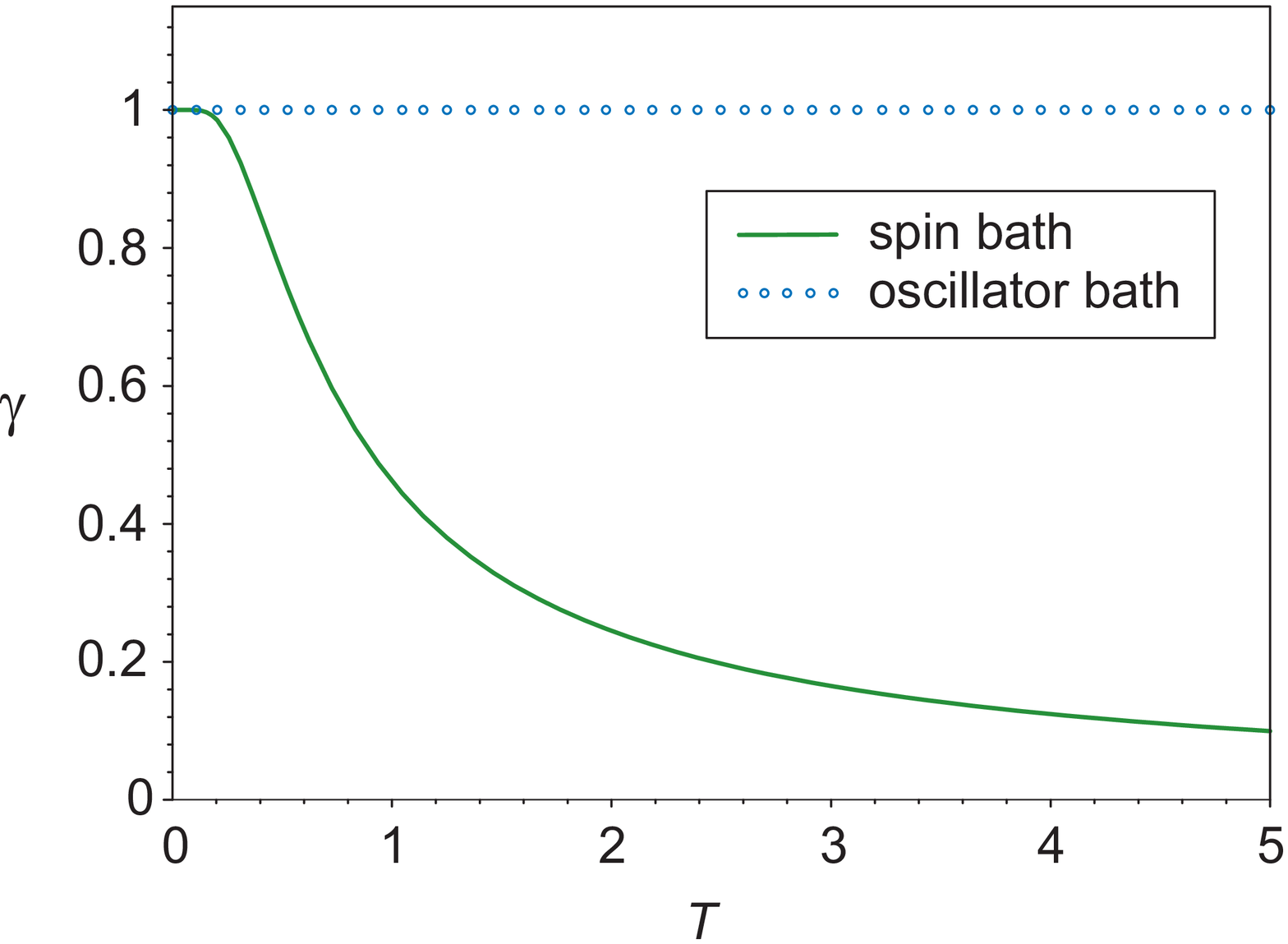}

\vspace{-0.2cm}

\begin{flushleft} \emph{(b)} \end{flushleft}

\vspace{-1cm}

\hspace{.01cm} \includegraphics[scale=.3]{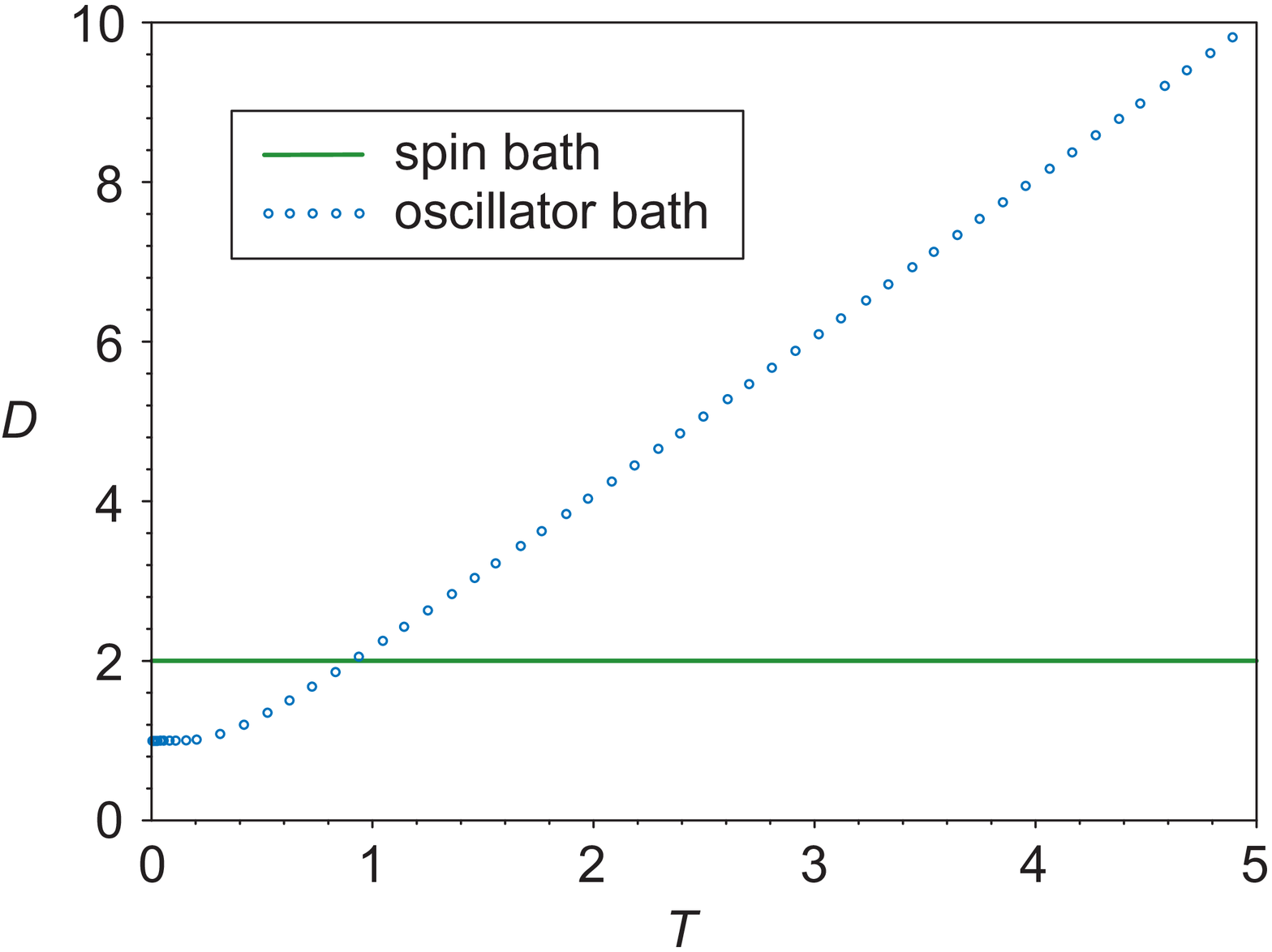}

\caption{\label{fig:coeff}Dissipation and normal-diffusion
  coefficients \emph{(a)} $\gamma$ and \emph{(b)} $D$ as a function of
  the bath temperature $T$ for the oscillator--spin model (solid line)
  and for quantum Brownian motion (circles), assuming the ohmic
  spectral density \eqref{eq:AApojsvsddsjldfv1}. The vertical axis is
  normalized in units of the zero-temperature values $\gamma(T=0)$ and
  $D_\text{QBM}(T=0)$, respectively.  The horizontal (temperature)
  axis is displayed in units of $\Omega_0/k_\text{B}$. We use $D_0 =
  D_\text{QBM}(T=0)$.}
\end{figure}

In Fig.~\ref{fig:coeff} we have plotted the temperature dependencies
of the coefficients $\gamma$ and $D$ for the spin bath [see
Eqs.~\eqref{eq:poisjuw48yuKKHG2} and \eqref{eq:pNUNjuw48yuKKHG2}] and
for an oscillator bath (quantum Brownian motion) [see
Eqs.~\eqref{eq:poisjuw48yu2} and \eqref{eq:poisjuw48yudfkskjbkl2}],
with both baths described by the spectral density
\eqref{eq:AApojsvsddsjldfv1}. Fig.~\ref{fig:coeff}a shows that the
spin-bath dissipation rate $\gamma$, Eq.~\eqref{eq:poisjuw48yuKKHG2},
decreases with temperature. This initially surprising result is easily
explained. While a harmonic oscillator can absorb an infinite amount
of energy, there are only two energy levels for a spin-$\frac{1}{2}$
particle. It follows that, as the bath temperature is raised, the spin
bath saturates quickly and the dissipative influence on the central
system must decrease when compared with that of the oscillator bath,
whose dissipation rate is temperature-independent (for linear quantum
Brownian motion). Indeed, the $\tanh \left(\Omega_0/2k_\text{B}
  T\right)$ temperature dependence of the spin-bath dissipation rate
\eqref{eq:poisjuw48yuKKHG2} has been explicitly observed in systems
such as glasses \cite{Golding:1976:un} where dissipation is mainly
caused by interactions between phonon modes and TLS
\cite{Phillips:1987:yb}.

A similar argument also allows us to understand the absence of any
temperature dependence of the normal-diffusion coefficient $D$ [see
Eq.~\eqref{eq:pNUNjuw48yuKKHG2}] for the spin bath and thus of the
rate of spatial decoherence
(Fig.~\ref{fig:coeff}b). Eq.~\eqref{eq:poisjuw48yudfkskjbkl2} shows
that for quantum Brownian motion this rate increases with temperature
as $\coth(\Omega_0/2k_\mathrm{B}T)$ (and linearly with $T$ in the
high-temperature limit of the Caldeira--Leggett model
\cite{Caldeira:1983:on}). This increase is due to the fact that, as
the temperature is raised, increasingly excited energy levels will be
occupied in each harmonic oscillator, and thus the characteristic
wavelengths present in the bath will decrease. Shorter environmental
wavelengths mean that the bath will be able to better resolve the
position of the central system, leading to stronger decoherence
(localization) of superpositions of well-separated positions. The
quick saturation of the spin bath with increasing temperature implies
that the characteristic wavelengths cannot become significantly
shorter, resulting for our model in a temperature-independent
expression for $D$.

The fact that the spin-bath decoherence rate $D$ has a constant
component implies that there is a heating source in the
oscillator--spin model. This term in the master equation drives a
diffusion process in the momentum and thus causes the average kinetic
energy of the oscillator to increase linearly in time at a rate of at
least $D_0$. This is likely to be a problematic source of heating in
micron-scale ion traps.  As this is independent of temperature, it
will be apparent even if no systematic dissipation is observable
($\gamma \approx 0$).

\section{\label{sec:oscill-coupl-single}Harmonic oscillator coupled to
  a single two-level system interacting with a bosonic bath}

The discussion in the previous section indicates that the temperature
dependence of the damping and dissipation rates arise from the fact
that the bath seen by the central oscillator cannot absorb an
arbitrarily large amount of energy. We are thus led to consider an
extreme case in which the oscillator is coupled to only a single TLS
which is itself weakly coupled to a bosonic bath. As mentioned in the
Introduction, both in micron-scale, cryogenic ion traps and in GHz
QEMS (the frequency regime relevant to the observation of quantum
effects) it is likely that only a few TLS take part in the dynamics,
rather than a TLS environment with continuous spectrum as used in
\cite{Seoanez:2006:yb,Seoanez:2007:um}. This lends particular
experimental relevance to the single-TLS model considered here. For
example, our model could represent a single charge trap
electrostatically coupled to a nanomechanical resonator, with the
charge trap itself coupled to Johnson--Nyquist electrical noise in the
surrounding circuit.

We begin with a single vibrational degree of freedom (the central
harmonic oscillator $\mathcal{S}$) coupled to a single
spin-$\frac{1}{2}$-particle (TLS),
\begin{equation}
  \op{H} = \hbar\Omega_0 \op{a}^\dagger \op{a} + g\hat{X} \otimes
  \op{\sigma}_z + \frac{\Delta}{2}\op{\sigma}_x, 
\end{equation}
where 
\begin{eqnarray}
  \left (\frac{\hbar}{2M\Omega_0}\right )^{-1/2}\hat{X}& = &
  (\op{a}+\op{a}^\dagger)=\hat{x},\\ 
  \left (2\hbar M\Omega_0\right )^{-1/2}\hat{P} & = & -i
  (\op{a}-\op{a}^\dagger) =\hat{p}, 
\end{eqnarray}
are dimensionless position and momentum operators.  In the case of a
realization of the model as a charge trap, $\op{\sigma}_z$ refers to
two distinct charge configurations of a single microscopic trap in one
of the electrodes near a nanomechanical resonator or trapped ion. We
could think of this as some kind of double well, in which case the
eigenstates of $\op{\sigma}_z$ refer to states localized on one side
of the barrier or the other. The tunnel-split ground states under the
barrier are eigenstates of $\op{\sigma}_x$ and the tunnel splitting is
$\Delta$. If we define the eigenstates of $\op{\sigma}_x$ by
$\op{\sigma}_x|\pm\rangle=\pm|\pm\rangle $, then we can define the
``bit-flip'' operator as $\op{\sigma}_z=|+\rangle\langle
-|+|-\rangle\langle +|$.

We expect that the TLS remains close to thermodynamic equilibrium with
a heat bath at temperature $T$ even in the presence of the coupling to
the central oscillator. Thus we assume that its state at all times can
be approximated by the thermal state
\begin{equation}
  \op{\rho}_{T}=p_+\ketbra{+}{+} + p_-\ketbra{-}{-},
\end{equation}
where 
\begin{equation}
\frac{p_+}{p_-}=\e^{-\Delta/k_\text{B} T}.
\label{TLS_state}
\end{equation}
The dynamical process that maintains the TLS in thermodynamic
equilibrium could be quite obscure. However, a simple model can be
given by weakly coupling the single TLS to a bosonic bath at
temperature $T$. The coupling is capacitive,
\begin{equation}
  \op{H}_{\text{coupling}}=\op{\sigma}_z \otimes \sum_kg_k\hat{q}_k(t),
\end{equation}
while the free Hamiltonian for the bath is a sum over harmonic
oscillators, each with canonical coordinates $\hat{q}_k,\hat{p}_k$ and
frequency $\omega_k$.  If the coupling is weak so that $g_k \ll
\Delta, \omega_k$, the corresponding Markov master equation for the
density operator $\op{\rho}$ of the joint oscillator--TLS system is
\cite{Walls:1994:zz}
\begin{align}
  \frac{\D\op{\rho}}{\D t} & = - \I \Omega_0\left[\op{a}^\dagger
    \op{a},\op{\rho}\right]- \I
  \Delta\left[\op{\sigma}_x,\op{\rho}\right]- \I
  g\bigl[\hat{X}\op{\sigma}_z,\op{\rho}\bigr] \notag \\ 
  & \quad + \gamma(\bar{n}+1){\cal
    D}[\op{\sigma}_-]\op{\rho}+\gamma\bar{n}{\cal
    D}[\op{\sigma}_+]\op{\rho},
    \label{ME}
\end{align}
where $\Omega_0$ is the vibrational frequency of the central
oscillator and $\gamma$ determines the heating rate.  The
super-operator ${\cal D}[\op{A}]$ is defined by
\begin{equation} {\cal D}[\op{A}]\op{\rho} \equiv \op{A}\op{\rho}
  \op{A}^\dagger-\frac{1}{2}(\op{A}^\dagger \op{A}\op{\rho}+\op{\rho}
  \op{A}^\dagger \op{A}),
\end{equation}
and $\op{\sigma}_+=\op{\sigma}_-^\dagger=|+\rangle\langle -|$ are
raising and lowering operators in the eigenstates of $\op{\sigma}_x$,
i.e., the energy eigenstates. Finally, the parameter $\bar{n}$ is
defined by
\begin{equation}
  \label{eq:nbar}
  \bar{n} \equiv \left (\e^{\beta\Delta/2}-1\right )^{-1}.
\end{equation}
It is easy to see that in the absence of the coupling to the
oscillator, the steady state for the TLS implied by Eq.~\eqref{ME} is
simply given by Eq.~\eqref{TLS_state}.

\begin{figure}

\begin{flushleft} \emph{(a) \,\, $\gamma=10$} \end{flushleft}

\vspace{-0.4cm}

\includegraphics[scale=.85]{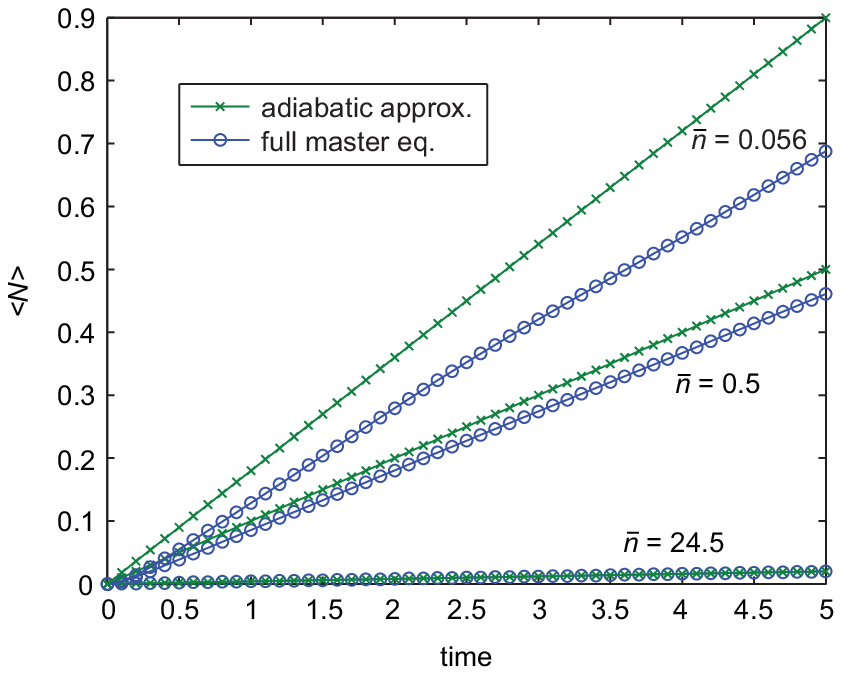}

\vspace{-0.4cm}

\begin{flushleft} \emph{(b) \,\, $\gamma=100$} \end{flushleft}

\vspace{-.1cm}

\includegraphics[scale=.85]{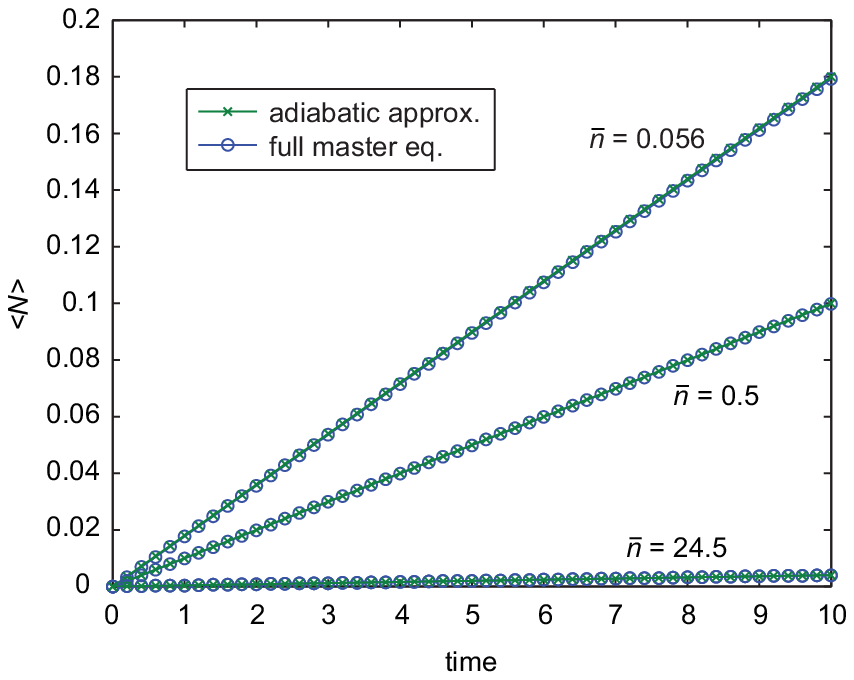}

  \caption{\label{fig:sim}Comparison of the adiabatic approximation
    (crosses) and the full master equation (circles) for the simple
    model discussed in Sec.~\ref{sec:oscill-coupl-single}. We show the
    mean occupation number $\langle N \rangle$ of the central harmonic
    oscillator as a function of time at three different temperatures
    (a larger value of $\bar{n}$, see Eq.~\eqref{eq:nbar}, corresponds
    to higher temperature). We use $\Delta=1$, $\Omega_0=1$, $g=1$,
    and \emph{(a)} $\gamma=10$, \emph{(b)} $\gamma=100$.}
\end{figure}

We can now calculate a heating rate for the oscillator. We do this by
adiabatic elimination \cite{Warszawski:2000:am} of the TLS, i.e., we
assume that $\gamma \gg \Omega_0,g,\Delta$ so that the TLS remains in
steady state slaved to the motion of the oscillator. We define the
operators acting on the vibrational degree of freedom by
$\op{\rho}_{++} \equiv \langle + |\op{\rho}|+\rangle,\ \
\op{\rho}_{--} \equiv \langle -|\op{\rho}|-\rangle,\ \ \op{\rho}_{+-}
\equiv \langle +|\op{\rho}|-\rangle$. Note that the reduced density
operator for the vibrational degree of freedom is just given by
$\op{\rho}_\mathcal{S} =\op{\rho}_{++}+\op{\rho}_{--}$.  The master
equation~\eqref{ME} then implies that
%
\begin{subequations}
\begin{align}
  \frac{\D\op{\rho}_{++}}{\D t} & =  -\I
  g(\hat{X}\op{\rho}_{-+}-\op{\rho}_{+-}\hat{X})-\I\Omega_0[\op{a}^\dagger
  \op{a},\op{\rho}_{++}] \notag \\ & \quad
  -\gamma(\bar{n}+1)\op{\rho}_{++}+\gamma\bar{n}\op{\rho}_{--},\\
  \frac{\D\op{\rho}_{--}}{\D t}  & =  -\I
  g(\hat{X}\op{\rho}_{+-}-\op{\rho}_{-+}\hat{X})-\I\Omega_0[\op{a}^\dagger
  \op{a},\op{\rho}_{--}] \notag \\ & \quad
  +\gamma(\bar{n}+1)\op{\rho}_{++}-\gamma\bar{n}\op{\rho}_{--},\\
  \frac{\D\op{\rho}_{+-}}{\D t} & =  -\I
  g(\hat{X}\op{\rho}_{--}-\op{\rho}_{++}\hat{X})-\I\Omega_0[\op{a}^\dagger
  \op{a},\op{\rho}_{+-}] \notag \\ & \quad
  -\frac{\gamma}{2}(2\bar{n}+1)\op{\rho}_{+-}-2\I\Delta\op{\rho}_{+-}.
\end{align}
\end{subequations}
%
Assuming that the off-diagonal operators $\op{\rho}_{+-}$ reach a steady
state, we find that
\begin{equation}
  \op{\rho}_{+-}\approx -\frac{2\I
    g}{\gamma(2\bar{n}+1)}(\hat{X}\op{\rho}_{--}-\op{\rho}_{++}\hat{X}). 
\end{equation}
Substituting this into the equation of motion for the diagonal
components leads to
\begin{equation}
  \label{eq:simmodelmeq}
  \frac{\D\op{\rho}_\mathcal{S}}{\D t}=-\I\Omega_0 \left[ \op{a}^\dagger
    \op{a},\op{\rho}_\mathcal{S} \right] - \Gamma \bigl[ \hat{X},
    \bigl[\hat{X},\op{\rho}_\mathcal{S}\bigr]\bigr],  
\end{equation}
where the last term implies diffusive heating and the rate is given by
\begin{equation}
  \Gamma=\frac{2g^2}{\gamma(2\bar{n}+1)}.
\end{equation}
A full numerical simulation (see Fig.~\ref{fig:sim}) confirms that
this master equation is a good description of the dynamics in the
limit of large $\gamma$. However, even when the adiabatic
approximation is not the same as the full master equation, the full
model still shows diffusive heating.

At zero temperature the TLS is in the ground state $|-\rangle$. As
this is a superpositon of the two eigenstates of $\op{\sigma}_z$, the
vibrational degree of freedom suffers momentum kicks of equal
magnitude but random sign. As the temperature goes to infinity, the
TLS state is the identity operator and the vibrational degree of
freedom suffers no kicks at all. For this reason the heating rate goes
to zero at high temperature---a reappearance of the damping-rate
feature of the oscillator--spin model discussed in
Sec.~\ref{sec:comp-betw-spin}. Also note that at zero temperature
there is a fixed momentum-diffusion rate that causes the oscillator to
heat, which again is the case for the full spin-bath model of
Sec.~\ref{sec:derivation}.
  
\section{\label{sec:conclusions}Summary and conclusions}

We have derived the master equations for a single harmonic oscillator
coupled (i) to a bath of two-level systems (the oscillator--spin
model) and (ii) to a single two-level system interacting with a bosonic
bath.  These models and the derivation of the relevant master
equations not only close an important gap in the set of canonical
system--environment models for decoherence and dissipation, but are
also motivated by and relevant to current experiments such as
quantum-electromechanical systems and micron-scale ion traps.

For both models the key features that arise are: (i) The systematic
dissipation rate for the oscillator {\em decreases} with increasing
temperature; (ii) at zero temperature the systematic dissipation rate
goes to zero; but (iii) there is a temperature-independent
momentum-diffusion rate (a heating rate).  Interestingly, this
behavior is very different from the model for quantum Brownian motion,
and we have explained how it can be understood as arising from a rapid
saturation of the spin environment.

An obvious direction for future investigations is the application of
our model to concrete experiments and a comparison of the theoretical
predictions with experimental data. In particular, we plan to carry
out direct simulations of the model in an ion trap.

\begin{acknowledgments}
  We thank Miles Blencowe, Ross McKenzie, and Philip Stamp for useful
  discussions. MS and GJM acknowledge support from the Australian
  Research Council. APH thanks the Pacific Institute of Theoretical
  Physics and PIMS Vancouver for support.
\end{acknowledgments}

\bibliographystyle{apsrev}


\end{document}